# Deep Radon Prior: A Fully Unsupervised Framework for Sparse-View CT Reconstruction

Shuo Xu, *Graduate Student Member, IEEE*, Yucheng Zhang, Gang Chen, Xincheng Xiang, Peng Cong, and Yuewen Sun

*Abstract*—Although sparse-view computed tomography (CT) has significantly reduced radiation dose, it also introduces severe artifacts which degrade the image quality. In recent years, deep learning-based methods for inverse problems have made remarkable progress and have become increasingly popular in CT reconstruction. However, most of these methods suffer several limitations: dependence on high-quality training data, weak interpretability, etc. In this study, we propose a fully unsupervised framework called Deep Radon Prior (DRP), inspired by Deep Image Prior (DIP), to address the aforementioned limitations. DRP introduces a neural network as an implicit prior into the iterative method, thereby realizing cross-domain gradient feedback. During the reconstruction process, the neural network is progressively optimized in multiple stages to narrow the solution space in radon domain for the under-constrained imaging protocol, and the convergence of the proposed method has been discussed in this work. Compared with the popular pre-trained method, the proposed framework requires no dataset and exhibits superior interpretability and generalization ability. The experimental results demonstrate that the proposed method can generate detailed images while effectively suppressing image artifacts. Meanwhile, DRP achieves comparable or better performance than the supervised methods.

## I. INTRODUCTION

As a non-invasive imaging method, X-ray computed tomography (CT) has been utilized in a variety of fields, including security inspections, clinical applications, and nondestructive tests (NDTs).

The potential harm caused by ionizing radiation has sparked growing concern regarding the minimization of CT dosages. One commonly employed technique is sparse-view CT, which involves a reduction in the number of projection views. Due to Nyquist sampling theorem, it is an extremely difficult task for traditional analytical and iterative CT reconstruction algorithms, such as filtered back-projection method(FBP), Expectation Maximization Algorithm (EM)[1], or simultaneous algebraic reconstruction technique(SART) [2], to get high-quality results from sparse sampling data. Sparse sampling leads to data incompleteness in the projection domain, which may result in severe noise and artifacts in the reconstructed images. In recent decades, more and more researchers are focusing on deep learning techniques driven by their great success in computer vision.

The reconstruction methods based on deep learning can be divided into two main approaches. **(1) Data restoration**. This type of method improves the quality of under-sampled measurement data by mapping projections and images separately into high-quality data. There are usually three ways for this method as following, image domain processing such as a deep convolutional neural network called FBPConvNet[3], projection domain processing such as a residual encoder-decoder CNN called RED-CNN[4], a denseNet deconvolution network called DD-Net[5], and hybrid domain processing like DuDoNet (Dual Domain Network)[6] and DRONE (Dual-Domain Residual-based Optimization Network)[7]. This type of algorithm can be easily deployed into the data flow of traditional algorithms and has certain advantages in terms of computation speed. Therefore, it has been widely researched in the fields of sparse view reconstruction, limited angle reconstruction and metal artifact removal. However, the consistency of the cross-domain progress between the original projection and the final image is not deliberated in this method, which leads to potential accuracy and robustness issues. **(2) End-to-end learning.** This approach takes the measurement values in radon domain as input and directly outputs the tomographic image, aiming to train a neural network to learn an inverse transformation operator corresponding to a specific imaging protocol, such as AUTOMAP[8] and iRadonMap[9]. However, this type of method is vulnerable to perturbation, which causes unstable results[10].

To improve the stability of the algorithm, an important branch is to combine neural networks with iterative algorithms. learned experts' assessment-based reconstruction network (LEARN)[11] replaced the regularization term with a neural network. ISTA-Net[12] and FISTA-Net[13] used a shrinkage/soft-threshold step to update the gradient. The learned primal-dual hybrid gradient algorithm accounts for a (possibly non-linear) forward operator in a deep neural network by unrolling a proximal primal-dual optimization method[14]. The approach of combining iterative algorithms with neural networks systematically overcomes the instability problem and has achieved excellent performance in some cases, but it still highly depends on data-driven training strategies[11-16].

Although neural network algorithms have higher performance top-limit than traditional algorithms under certain circumstances, they also have several limitations and shortcomings.

(1) Data dependence: Neural networks usually require larger data sets, often consisting of at least thousands or even millions of labeled samples, compared to traditional algorithms. Additionally, the generalization performance of algorithms based on deep learning is weak. Especially for medical imaging, the data sets are highly specialized, with different modes and concerned organ classification leading to significant differences in images. When the statistical properties of the input data are significantly different from the training set image, neural network algorithms may produce unpredictable errors.

(2) Data ethics: In the field of CT reconstruction, high-quality labeled images are mostly obtained through high-dose CT scanning, which exposes patients to high levels of radiation. Hence, the ethics and governance surrounding data sources are currently under dispute.

(3) Weak interpretability: One of the most notable shortcomings of neural networks is their "black box" nature, which hinders our ability to understand how and why certain outputs are produced. For instance, when the input projection data is fed into the network once, the reconstruction results may deteriorate in some areas and produce undesired artifacts. It is a daunting task to explain the root cause of such errors. By incorporating interpretable features, however, it is easier to comprehend the reasons behind the errors.

(4) Immaturity of unsupervised algorithms: Recently, some Deep Image Prior (DIP) related unsupervised algorithms have been proposed[17-19]. However, these algorithms only act on the image domain and lack consistency constraints on cross-domain information. The performance of the algorithm is limited and remains less than satisfactory. In addition, some other algorithms that purport to be unsupervised are still in a semi-supervised stage, requiring pre-trained models or data-driven prior knowledge.

The aforementioned shortcomings and limitations have resulted in the lack of robustness in neural network-related and DIP-related CT reconstruction algorithms. However, the outstanding fitting ability of neural networks provides this mature system with great potential and versatility. Therefore, we explore an innovative unsupervised framework in this paper, called deep radon prior (DRP), which introduces deep learning into iterative reconstruction algorithm. We derive an equivalent transformation, embed optimization target into the neural network, utilized the projection residual under radon transformation as the target, and optimize the network and image parameters simultaneously. This method is more formally similar to the model-based iterative algorithm. As an auxiliary fitting function, the neural network acts as a hyperparametric iterative operator. This method expands the dimension and scope of the solution domain, thereby improving the upper limit of accuracy. It can also be understood as a general constraint that optimizes the image through information perception in radon domain.

Our main contributions can be summarized as follows:

(1) An innovative radon domain loss function is proposed for CT reconstruction problem, which extracts implicit prior in radon domain and enables cross-domain gradient feedback.

(2) A novel fully unsupervised framework for ill-posed inverse problem is established, which combines neural networks with traditional iterative reconstruction algorithms. In the field of sparse-view reconstruction, this method outperforms supervised neural network-based methods, while exhibiting strong generalization and robustness.

(3) The convergence of the proposed optimization scheme is proved in this paper, and the differences between the proposed method and traditional algorithms in iterative progress are analyzed.

The remainder of this paper is organized as follows. In Section II, CT reconstruction algorithms and the related work are introduced in detail. Section III presents the proposed framework. In Section IV, experimental studies and results are demonstrated, including comparisons with traditional algorithms and several famous advanced algorithms. Then, the corresponding discussion is in Section V. Finally, the conclusion is given in Sections VI.

## II. RELATED WORK

### A. Classical Reconstruction Algorithm

In general, traditional CT reconstruction algorithm can be classified into two categories: analytical algorithm and iterative algorithm. The analytical algorithm uses inverse radon transform derived from the central slice theorem, and is the earliest studied CT reconstruction algorithm. At present, the well-known algorithms such as Filtered back-projection (FBP), Back-projection Filtration (BPF)[20] and other algorithms[21-23] are commonly used. However, the ideal assumption of the analytical algorithm is contrary to the discrete sampling of sparse angles. Therefore, in this case, the reconstructed image will suffer from strong steak artifacts, rendering the analytical algorithm unsuitable for solving the problem in this study. In addition, the analytical algorithm uses a different set of mathematical languages from the iterative algorithm, which we will not introduce in this paper. This section focuses on the iterative algorithm.

Typically, the CT reconstruction problem is treated as solving a linear Equation:

$$p = Ax = \begin{bmatrix} \alpha(\theta_1,\upsilon_1) & \alpha(\theta_2,\upsilon_2) & \cdots & \alpha(\theta_1,\upsilon_N) \\ \alpha(\theta_2,\upsilon_1) & \alpha(\theta_2,\upsilon_2) & \cdots & \alpha(\theta_2,\upsilon_N) \\ \vdots & \vdots & \ddots & \vdots \\ \alpha(\theta_{M\cdot W},\upsilon_1) & \alpha(\theta_{M\cdot W},\upsilon_2) & \cdots & \alpha(\theta_{M\cdot W},\upsilon_N) \end{bmatrix} \cdot \begin{bmatrix} x_1 \\ x_2 \\ \vdots \\ x_N \end{bmatrix} \quad (1)$$

Where $x = [x_1 \cdots x_N]^T$ denotes a vector of discrete attenuation coefficients for a tomography image, $p = [p_1 \cdots p_{M\cdot W}]^T$ represents the measured data after calibration and log-transform, $A \in \mathbb{R}^{(M\cdot W)\times N}$ is the imaging system or projection matrix corresponding to a specific configuration of the CT system with angular sampling number M (In the field of sparse-view imaging, M is usually less than or equal to 180 with uniform angular spacing), detector unit number W and image pixel or voxel number N, and $\alpha(\theta_{M \cdot W}, \upsilon_N)$ indicates the forward coefficient from different pixel or voxel location $\upsilon$ to a sinogram object θ. The purpose of image reconstruction is to recover the unknown x from the system matrix A and observed data p.

For iterative image reconstruction, equation (1) can be solved by minimizing the following constrained objective function:

$$\begin{aligned} x &= \arg\min_x \xi(x) \\ &= \arg\min_x \frac{\lambda}{2}\|Ax - p\|_\omega^2 + R(x), \; s.t. \; x_j \geq 0 \; \forall j \end{aligned} \quad (2)$$

Where $\|\cdot\|_\omega^2$ represents a certain norm, which is usually L2 norm. This term is for data fidelity, which constraints the consistency between image $x$ and observed radon domain data P. the second term $R(x)$ is for regularization, and $\lambda$ controls the balance between data fidelity and regularization.

When $R(x) = 0$, this method has been widely used in previous studies, such as the algorithms like SART, EM, etc. The effect of these algorithms is acceptable when the projected image quality is enough. However, in most cases, we need to add R(x) as a regularization constraint to further improve the image quality, especially for ill-posed inverse problem such as sparse-view CT reconstruction. In the past few decades, scholars have explored many kinds of regularization term designs, such as the famous Total Variational (TV)[24, 25] and its various variants, including total generalized variation (TGV)[26, 27], nonlocal TV (NLTV)[28], reweighted anisotropic total variation[29] and anisotropic relative total variation[30]. Regularization terms are based on artificial design. Unfortunately, in the field of sparse-view reconstruction, the performance of traditional iterative algorithms is still limited.

*B. Neural Network-Related Methods*

Recently, deep learning has achieved remarkable success in the field of image processing and computer vision, providing a powerful tool for solving imaging problems. Therefore, researchers have integrated deep learning technology into CT reconstruction, which can be broadly categorized into three groups: image domain processing, projection domain processing, and deep reconstruction methods.

Image domain processing methods construct a neural network to post-process the image reconstructed by the traditional algorithm, which can be described as the following expression:

$$\begin{aligned} \hat{x} &= \mathcal{O}_{IDP}(x; \vartheta_{IDP}) \\ \vartheta_{IDP} &= \arg\min_{\vartheta_{IDP}} L(\hat{x}, x^\dagger)_{train} \end{aligned} \quad (3)$$

Where $\mathcal{O}_{IDP}(\cdot\,; \vartheta_{IDP})$ denotes the neural-network model parameterized by $\vartheta_{IDP}$, $x^\dagger$ represents the ground truth (without any noise or artifacts), and $L(\hat{x}, x^\dagger)$ is the loss function between $\hat{x}$ and $x^\dagger$.

For projection domain processing, the problem can be solved by inpainting sinogram:

$$\begin{aligned} \hat{p} &= \mathcal{O}_{PDP}(p; \vartheta_{PDP}) \\ \vartheta_{PDP} &= \arg\min_{\vartheta_{PDP}} L(f_{PDP}(\hat{p}), x^\dagger)_{train} \end{aligned} \quad (4)$$

Where $\mathcal{O}_{PDP}(\cdot\,; \vartheta_{PDP})$ denotes the neural-network model parameterized by $\vartheta_{PDP}$, $p \in \mathbb{R}^{(M \cdot W) \times 1}$ represents the measured projection data after calibration and log-transform. $\hat{p} \in \mathbb{R}^{(M_{PDP} \cdot W) \times 1}$ is the sinogram with enough angular sampling number of $M_{PDP}$ (usually greater than or equal to 360), and $f_{PDP}(\cdot)$ is a forward operator that can usually be replaced by algorithms such as FBP. Projection domain processing is a feasible idea and is usually not used independently. At present, the algorithm based on the dual-domain hybrid backbone is a popular research direction.

For deep reconstruction methods, there is no unified technical solution as it remains an open problem. There have been some notable works in this area, including LEARN, Parameterized Plug-and-Play ADMM (3pADMM), CNN-Based Projected Gradient Descent[31, 32], etc. There is no unified formula for this method.

In general, the above neural network-based methods are still not suitable for large-scale deployment and implementation, as described in Section I.

*B. Deep Image Prior (DIP)-Related Methods*

Deep Image Prior (DIP) is a revolutionary unsupervised algorithm proposed in 2018 for image restoration, which means recovering unknown real images from their inferior images, including image denoising, super-resolution and image inpainting. These tasks are similar to the challenges faced in the field of CT reconstruction, making DIP an instructive method for this area. Therefore, the proposed method in this paper is strongly inspired by DIP from the design rationale.

DIP can be described as the following expression:

$$\begin{aligned} x &= \mathcal{O}(z; \vartheta^*) \\ \vartheta^* &= \arg\min_\vartheta L(\mathcal{O}(z; \vartheta), x_0) \end{aligned} \quad (5)$$

Where $\mathcal{O}(\cdot\,; \vartheta)$ denotes the neural-network model parameterized by $\vartheta$, The minimizer $\vartheta^*$ is obtained using an optimizer such as gradient descent starting from a random initialization of the parameters, $x_0$ is the noisy/low-resolution/occluded image, $z$ represents a random code vector. This approach can be used to sample realistic images from a random distribution. Since no aspect of the network is pre-trained from data, this approach is effectively handcrafted, just like the TV norm. The experiments demonstrate that this hand-crafted prior works well for various standard inverse problems such as denoising, super-resolution, and inpainting. Moreover, it bridges the gap between two popular families of image restoration methods: learning-based methods and learning-free methods based on handcrafted image priors.

To minimize the reconstruction error, an untrained network can be trained during the reconstruction process[19], which can be expressed as:

$$\begin{aligned} x &= \mathcal{O}(z; \vartheta^*) \\ \vartheta^* &= \arg\min_\vartheta \xi(\vartheta) = \arg\min_\vartheta \|A\mathcal{O}(z; \vartheta) - p\|_2 \end{aligned} \quad (6)$$

Where Shu et al. used the normal operator of matrix A to accelerate the computation and transformed the equation into:

$$x = \mathcal{O}(z^*; \vartheta^*)$$
$$z^*, \vartheta^* = \arg\min_{z,\vartheta} \|A^T \cdot A\mathcal{O}(z,\vartheta) - A^T \cdot p\|_2 \quad (7)$$

In (7), $A^T \cdot p$ is the back-projection result of the detected sinogram $p$, which can be regarded as an approximation of the reconstruction images, and $A^T \cdot A$ can be implemented with a convolution kernel[33, 34]. Thus, (7) can be considered as a variant framework of deep image prior, which is abbreviated as UN-DIP and has been proven to be effective in CT reconstruction problem. However, the introduction of $A^T$ exacerbates the optimization challenges and computational costs, thereby impeding the performance of these methods.

## III. PROPOSED METHOD

To maintain the generalizability of our approach, we formulate sparse-view CT reconstruction problem under a parallel beam system configuration. Other systems such as equiangular fan beam configuration, equidistance fan beam configuration and linear CT system are also equivalent under data rebinning[35]. Another advantage of the parallel beam system is that the geometric arrangement of DSO (Distance between Source and Origin) and DSD (Distance between Source and Detector) has no effect on the reconstruction results. Therefore, this configuration is an ideal experimental design, which minimizes the disturbances outside the reconstruction algorithm.

*A. Design Rationale*

*1) Optimization objective*

TABLE I
COMPARISON OF THE OBJECTIVES FOR DIFFERENT METHODS

| Method | | Optimization objectives |
|---|---|---|
| Classical Method | Iterative algorithm | $x = \arg\min_x \xi(x) = \arg\min_x \frac{\lambda}{2}\|Ax-p\|_\omega^2 + R(x)$ |
| Neural Network-Related Method | Image Domain Processing | $\hat{x} = \mathcal{O}_{IDP}(x; \vartheta_{IDP})$ <br> $\vartheta_{IDP} = \arg\min_{\vartheta_{IDP}} L(\hat{x}, x^\dagger)_{train}$ |
| | Projection Domain processing | $\hat{p} = \mathcal{O}_{PDP}(p; \vartheta_{PDP})$ <br> $\vartheta_{PDP} = \arg\min_{\vartheta_{PDP}} L(f_{PDP}(\hat{p}), x^\dagger)_{train}$ |
| DIP-Related Method | Deep Image Prior | $x = \mathcal{O}(z; \vartheta^*)$ <br> $\vartheta^* = \arg\min_\vartheta L(\mathcal{O}(z;\vartheta), x_0)$ |
| | UN-DIP | $x = \mathcal{O}(z^*; \vartheta^*)$ <br> $z^*, \vartheta^* = \arg\min_{z,\vartheta} \|A^T \cdot A\mathcal{O}(z,\vartheta) - A^T \cdot p\|_2$ |
| Proposed Method | Deep Radon Prior | $x = \mathcal{O}(z^*; \vartheta^*)$ <br> $z^*, \vartheta^* = \arg\min_{z,\vartheta} \xi(z,\vartheta) = \arg\min_{z,\vartheta} L(A\mathcal{O}(z;\vartheta), p)$ |

Inspired by DIP and the aforementioned work, we have reevaluated (2). We replace the regularization term R(x) with the implicit prior captured by the neural network. This is an equivalent transformation that expands the breadth of the solution domain:

$$x = \mathcal{O}(z^*; \vartheta^*)$$
$$z^*, \vartheta^* = \arg\min_{z,\vartheta} \xi(z,\vartheta) = \arg\min_{z,\vartheta} L(A\mathcal{O}(z;\vartheta), p) \quad (8)$$

Where $\mathcal{O}(\cdot;\vartheta)$ denotes the neural-network model parameterized by $\vartheta$, $x \in \mathbb{R}^{N^2 \times 1}$ is the tomography image that needs to be reconstructed, $z \in \mathbb{R}^{N^2 \times 1}$ represents a random code vector. z has the same spatial size as $x$, hence it can also be considered as a temporary state during the iterative progress.

To prove the rationality of this method, we compared the proposed equation with several other schemes:

(I) Equation (6): The solution range of (8) is larger, which conforms to the breadth-first search strategy. When solving (6), due to the fact that the input z has not been optimized, the algorithm may lose its exploration ability when it is not close to the true optimal solution, which can result in the algorithm getting stuck in a local optimal solution and failing to achieve global optimization.

(II) Equation (7): In fact, it is a heuristic strategy of (8). Shu et al. believe that it is extremely difficult to solve (8) due to the ill-conditioning and sparsity of A. Therefore, they propose (7) to obtain an approximate solution, which is not guaranteed to be the optimal solution and may even lead to the possibility of getting a spurious solution.

(III) Equation (2): The optimization objective of the iterative algorithm is actually equivalent to (8). $A\mathcal{O}(z;\vartheta) = p$ is a domain transformation of $Ax = p$ by setting $x = \mathcal{O}(z;\vartheta)$. From this perspective, the network model $\mathcal{O}(\cdot;\vartheta)$ is a parameterized iterative operator. However, unlike ordinary iterative operators, the neural network also plays a role in constraining and modifying the image due to its frequency properties. As the parametrization is known to present high impedance to image noise and unnatural texture, which has been proved by DIP, we have a reasonable prospect that it can be used to filter out noise and reduce the artifacts.

(IV) Equation (5): It should be emphasized that our method is not aimed at recovering data from image but rather recovering image from data. Our algorithm also uses image encoding as the initial value, but it applies constraints in the projection domain through radon transform, which is fundamentally difference from DIP.

*2) Optimization scheme*

Through the above analysis, we can conclude that (8) is a reasonable design, and its solution is legitimate and universal. However, an obvious obstacle is that this (8) is difficult to solve, which is also a factor that other researchers usually avoid discussing. To solve (8), we design an optimization scheme of alternating gradient descent, which is inspired by ADMM. The reconstruction process of DRP is shown in Algorithm.1.

---

**Algorithm.1** Deep Radon Prior (DRP)

**Input:** measure matrix A, measurement projection p, input z (FBP output image at the beginning), and step size β.

**Output:** reconstructed tomography image $x$

1: **Initialize:** $\Delta$, $\xi$, $\vartheta$, $\mathcal{O}(\cdot, \vartheta)$

2: $z = f_{FBP}(p)$

3: $\Delta = A^T \cdot (p - A\mathcal{O}(z;\vartheta))$ // gradient of image

4: $\xi = L(A\mathcal{O}(z;\vartheta), p)$ // Loss function of projection domain

5: **Repeat:**

6: $z = \mathcal{O}(z;\vartheta) + \beta \cdot \Delta$ // basic gradient descent

7: $\xi = L(A\mathcal{O}(z;\vartheta), p)$

8: $\vartheta = \arg\min_{\vartheta} \xi(\vartheta)$ // optimized by SGD or ADAM, etc.

9: $\Delta = A^T \cdot (p - A\mathcal{O}(z;\vartheta))$

10: **Until:** convergence, or some fixed number of iterations is reached

11: **Return:** $x = \mathcal{O}(z^*; \vartheta^*)$ // $\vartheta^*$ is the optimized parameters

---

In the Algorithm.1, $f_{FBP}(p)$ denoted the filtered back-projection algorithm. $A^T$ can also be replaced by back-projection operator. In Line 1, $\mathcal{O}(\cdot;\vartheta)$ corresponds to a model with parameter $\vartheta$. In this paper, the neural network is used as $\mathcal{O}(\cdot;\vartheta)$, but it can be replaced by other models, depending on the research achievements in the field of deep learning and optimization algorithms. As a framework, DRP is not limited to specific models and is open to future techniques. Lines 2 correspond to image initialization. In this step, we use the reconstruction results of FBP and find that it is feasible. Lines 3-10 correspond to the gradient descent algorithm. In the alternating gradient descent process, line 7 and line 8 is obviously reasonable because it is the optimization scheme of all deep learning-based algorithms. While the design of line 6 will be demonstrated in Section III-A-3.

In principle, the proposed framework realizes cross-domain gradient feedback through a neural network, and also closely links the iterative algorithm with unsupervised learning.

*3) Convergence discussion*

In this section, we will discuss the convergence of the proposed optimization strategy shown in algorithm.1. Without losing generality, we assume the loss function in the proof is L2 norm uniformly, which is a widely recognized scheme in image reconstruction. Many other norms are not derivable, which brings some difficulties to obtain a strict proof. However, the significance of more rigorous proof has exceeded the practical value of this article, so we present a simple form.

Our aim is to derive a proposition as the following:

**Proposition 1:** Let $\mathcal{O}(z_i;\vartheta_i)$ and $\mathcal{O}(z_{i+1};\vartheta_{i+1})$ be the two continuous solutions in the i$^{th}$ iteration process of (8). Where $\mathcal{O}(z_{i+1};\vartheta_{i+1})$ is obtained by the following as the Algorithm.1:

$$z_{i+1} = \mathcal{O}(z_i;\vartheta_i) + \beta \cdot \Delta$$
$$\vartheta_{i+1} = \arg\min_{\vartheta} \|A\mathcal{O}(z_{i+1};\vartheta_i) - p\| \quad (9)$$

Where:

$$\Delta = A^T \cdot (p - A\mathcal{O}(z_i;\vartheta_i)) \quad (10)$$

Then the following hold:

$$\|A\mathcal{O}(z_{i+1};\vartheta_{i+1}) - p\|_2^2 \le \|A\mathcal{O}(z_i;\vartheta_i) - p\|_2^2 \quad (11)$$

The crucial part of the proof is to construct a set of parameters for equivariant transfer as is shown in the following Lemma 1.

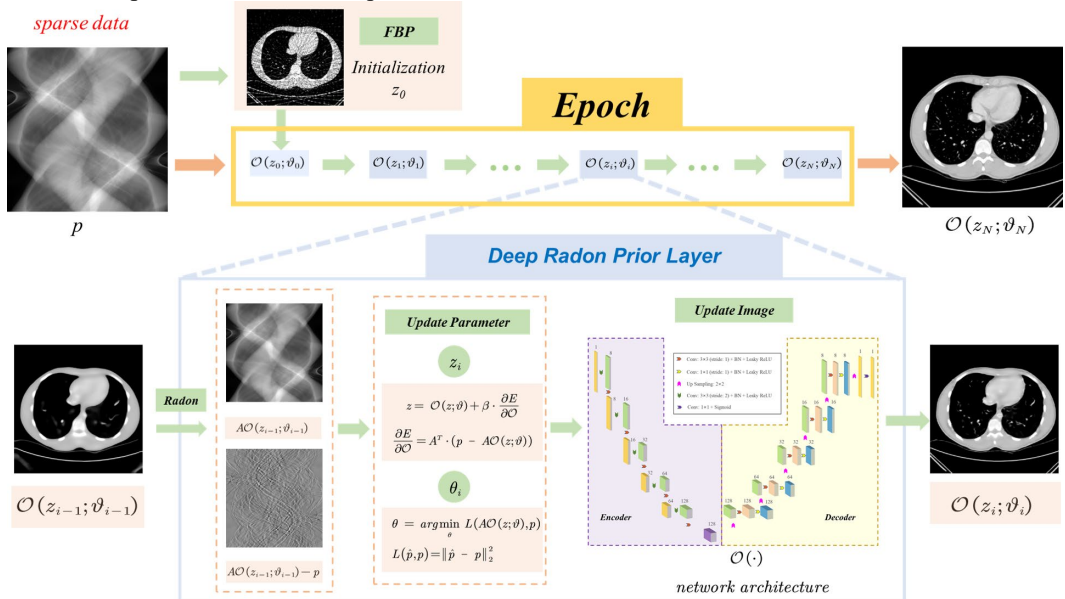

**Fig. 1.** Diagram of Deep Radon Prior.

**Lemma 1**: Suppose an image processing neural network $\mathcal{O}(\cdot;\vartheta)$ is designed with sufficient parameters and a reasonable structure, then the following hold: There always exist a set of parameters $\vartheta_{eq}$ that make it an equivariant function:

$$\exists \vartheta_{eq}, \ s.t. \ \mathcal{O}(z;\vartheta_{eq}) = z, \ \forall z \in \mathbb{R}^{N^2 \times 1} \tag{12}$$

*Proof of Lemma 1:* The result of Lemma 1 is not surprising, and it is obvious in the field of deep learning. Given the powerful expression capability of neural networks, it is painless to train a set of equivalent transformation parameters, which is simpler than tasks such as denoising and super-resolution. Since this is an obvious lemma, we only provide a literal sketch of the proof. More evidence can be found in the references[36-39].

**Lemma 2**: Assume Lemma 1 holds, then:

$$\|A\mathcal{O}(z;\vartheta^*) - p\|_2^2 \leq \|Az - p\|_2^2 \tag{13}$$

Where: $\vartheta^* = \arg\min_\vartheta \|A\mathcal{O}(z;\vartheta) - p\|_2^2$, $\mathcal{O}(z;\vartheta^*)$, $z \in \mathbb{R}^{N^2 \times 1}$

*Proof of lemma2:*
By lemma 1, we can obtain:

$$\|A\mathcal{O}(z;\vartheta^*) - p\|_2^2 \leq \|A\mathcal{O}(z;\vartheta_{eq}) - p\|_2^2 = \|Az - p\|_2^2 \tag{14}$$

Actually, it is not necessary to guarantee that $\vartheta^*$ is the global optimal solution. Instead, an ingenious scheme is to use $\vartheta_{eq}$ as the initial value. Even if the optimization only converges to a local optimum, the resulting solution will still perform better than $\vartheta_{eq}$. Therefore, (14) is strictly established. On the other hand, because our optimization goal is only one set of data, which is different from traditional supervised learning, the parameter search has become a simpler and more efficient problem. In experiments, we found that even if $\vartheta_{eq}$ is not strictly used as the initial value of each epoch, the result remains almost unchanged. For the further explanation of this operation, we shall give a point of view in the iterative process analysis in Section IV.

**Lemma 3**:

$$\|Az' - p\|_2^2 \leq \|Az - p\|_2^2 \tag{15}$$

Where $z' = z + \beta \cdot A^T(p - Az)$

*Proof of lemma 3:* This is the approach of Gradient-Descent [40]. We provide a simple proof under the situation using L2 norm. The proof is established by calculating the gradient:

$$\frac{\partial \|Az - p\|_2^2}{\partial z} = \frac{\partial (Az - p, Az - p)}{\partial z}$$

$$= \frac{\partial ((Az, Az) - 2(Az, p))}{\partial z} \tag{16}$$

$$= 2A^T(Az - p) = 2\Delta$$

Thus $\Delta$ is the gradient direction of $\|Az - p\|_2^2$. Then we can obtain:

$$z' = z - \beta_0 \frac{\partial \|Az - p\|_2^2}{\partial z} = z + 2\beta_0 A^T(p - Az) \tag{17}$$

$$= z + \beta A^T(p - Az)$$

Where: $\beta = 2\beta_0$. This is to make the form more graceful and for notational simplicity. According to the gradient descent method, $z'$ is closer to the optimal solution $z^*$. Hence, from the mathematically expectation, we can expect that: $\|Az' - p\|_2^2 \leq \|Az - p\|_2^2$.

From the discussion of the above three lemmas, it is sufficient to prove (11).

*Proof of Proposition 1:* By (9), (13), and (15), we shall estimate the boundary:

$$\|A\mathcal{O}(z_{i+1};\vartheta_{i+1}) - p\|_2^2$$
$$\leq \|Az - p\|_2^2 = \|A\mathcal{O}(z_i;\vartheta_i) + \beta \cdot A^T(p - Az) - p\|_2^2 \tag{18}$$

Consider $\mathcal{O}(z_i;\vartheta_i)$ as a whole item, then $A^T(p - Az) = \frac{\partial \|A\mathcal{O} - p\|_2^2}{\partial \mathcal{O}}$ is the gradient direction. Hence, further contraction:

$$\|A\mathcal{O}(z_i;\vartheta_i) + \beta \cdot A^T(p - Az) - p\|_2^2 \leq \|A\mathcal{O}(z_i;\vartheta_i) - p\|_2^2 \tag{19}$$

Combining (21) and (22):

$$\|A\mathcal{O}(z_{i+1};\vartheta_{i+1}) - p\|_2^2 \leq \|A\mathcal{O}(z_i;\vartheta_i) - p\|_2^2 \tag{20}$$

Q.E.D.

*B. Implementation Details*

In this study, the reconstruction resolution is set to 512 × 512, and the detector number is 724, which is larger than the diagonal length of image. The radon transform in the neural network is implemented by Torch-Radon[41] and Pytorch library. The loss function is optimized using the Adam method[42], and the learning rate is set to $5 \times 10^{-4}$, while the convolution kernels are initialized according to the random Gaussian distribution with zero mean and a standard deviation of 0.01. The image is reconstructed after 250 epochs, with the set of parameters $\vartheta$ of the neural network being optimized by 200 iterations during each epoch. The computations have been performed on one PC with an i5-10400 CPU, 16 GB of RAM, and an RTX 3060 GPU using Python. In Algorithm.1, the step size $\beta$ is set to 0.25 and we use the L2 norm for loss function.

$$L(A\mathcal{O}(z;\vartheta), p) = \|A\mathcal{O}(z;\vartheta) - p\|_2^2 \tag{21}$$

*C. Network Architecture*

In proposed DRP, the input and output of the neural network are tomography images, while the gradient is fed back from radon domain, which includes a non-local transformation. Thus, the DRP desires non-local neural networks. Similar with deep image prior (DIP), the proposed DRP restores images through training a randomly-initialized network. Therefore, as shown in Fig. 2, this paper selects the encoder-decoder network used in DIP as the backbone of the network. The network architecture has been proven to perform well in many tasks such as noise reduction, super-resolution, segmentation, recognition.

However, it should be emphasized that the proposed network in this paper has achieved the effect shown in Section IV without network architecture search and artificial optimization. In actual deployment, for the data generated by specific devices, network architecture search with pruning, distillation and other technologies can be used to further optimize the network design, improving the imaging performance of the DRP framework[43].

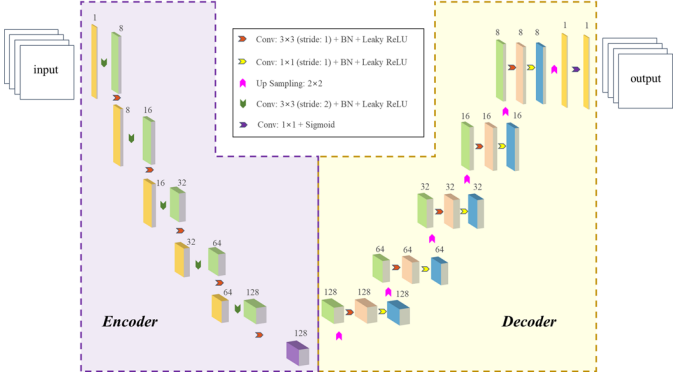

**Fig. 2.** A feasible network architecture for Deep Radon Prior. Abbreviations: Conv, convolution; BN, batch normalization; ReLU, rectified linear unit.

## IV. EXPERIMENTS AND RESULTS

In the following analysis, we use the structural similarity index (SSIM), and peak signal-to-noise ratio (PSNR) between the reconstructed image $x$ and ground truth $x^*$ for quantitative evaluation.

### A. Data Resources

To evaluate the imaging performance of the DRP framework under realistic conditions, two sets of open source CT datasets have been used in this paper, which are "CHAOS - Combined (CT-MR) Healthy Abdominal Organ Segmentation" [44] and "A Large-Scale CT and PET/CT Dataset for Lung Cancer Diagnosis (Lung-PET-CT-Dx)"[45]. Each data set of CHAOS consists of 16-bit DICOM images with a resolution of 512×512, x-y spacing between 0.7-0.8 mm and an inter-slice distance (ISD) of 3 to 3.2 mm. The "Lung-PET-CT-Dx" dataset consists of CT and PET-CT DICOM images of lung cancer subjects. The CT resolution was 512×512 pixels at 1mm×1mm, with a slice thickness and an interslice distance of 1mm.

Though DRP is an unsupervised algorithm that does not require the support of big data, we still need images to train supervised networks for comparison. In addition, the test images used in this study are also from the above two datasets.

### B. Iteration Process

Before comparing the methods, we visually analyze the iteration process of DRP algorithm, which is conducive to understand the effectiveness of the framework. Fig. 4 shows the DRP reconstruction results for different numbers of iterations, while Fig. 3 shows the results using traditional ADMM method with the TV norm for different numbers of iterations. The comparison reveals that the two frameworks adopt reverse update paths in the iterative process. In the first iteration, DRP only reconstructed the halo-like contour. With further iterations, DRP produces a large-scale, smooth and closed shape, forming a blurred image. On this basis, DRP gradually restores the details of the image, which is very similar to the characteristics of DIP. Excluding the incipient halo-like state, we can regard the iterative process of DRP after dozens of epochs as an addition in terms of information entropy. Starting from the most natural part of the image, DRP gradually adds disordered information. On the other hand, the traditional ADMM method is a subtraction progress when using FBP reconstructed image as the initial value. From the chaotic system with artifacts, ADMM gradually reduce the amount of information, so as to approach the original image.

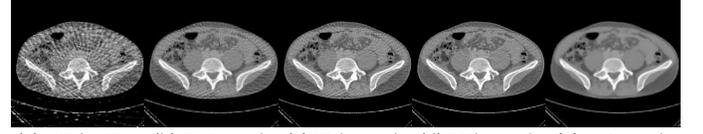

(a)Initialization  (b) First epoch  (c) 10th epoch  (d) 20th epoch  (e) Last epoch

**Fig. 3.** The reconstruction image of ADMM-TV algorithm at different iteration. The scanning view of the CT system is set to be 60.

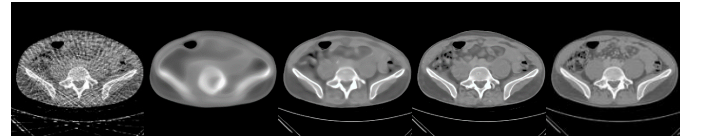

(a)Initialization  (b) First epoch  (c) 10th epoch  (d) 20th epoch  (e) Last epoch

**Fig. 4.** The reconstruction image of Deep Radon Prior algorithm at different iteration. The scanning view of the CT system is set to be 60.

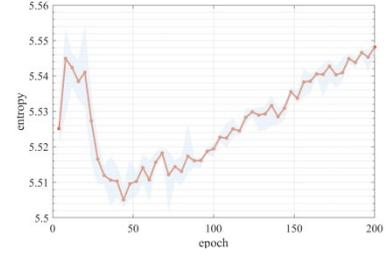

**Fig. 5.** The typical entropy curve of Deep Radon Prior algorithm. The scanning view is set to be 30. The aquamarine area is the deviation of the entropy.

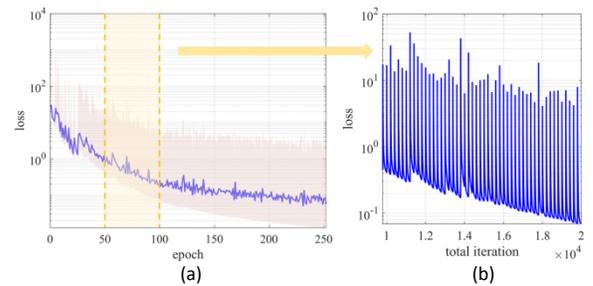

(a)                                          (b)

**Fig. 6.** The typical reconstruction loss curve of DRP algorithm under 30-views configuration. The pale pink area is the uncertainty of the loss function. The right figure shows the indicators of all iterations in 50-100 epoch.

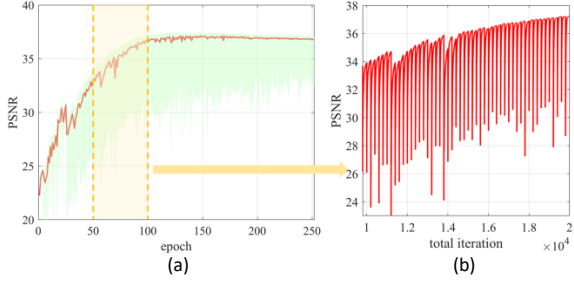

**Fig. 7.** The typical reconstruction PSNR curve of DRP algorithm under 30-views configuration. The aqua area is the uncertainty of the PSNR. The right figure shows the indicators of all iterations in 50-100 epoch.

To be precise, the information entropy curve of the DRP framework undergoes a decrease followed by an increase, as illustrated in Fig. 5. During the initial several dozen epochs, DRP generates several disordered and chaotic images. Such images contain many possibilities, resulting in a high entropy level, which is not the desired outcome. The objective is to produce is an image with ordered and precise information. To some extent, the essential characteristic of DRP is to transition numerous possibilities into detailed information, where the usual way is from high chaos to stability and subsequently to order.

The principle of entropy increase is a fundamental truth in nature. In an isolated system, entropy increase occurs easily and irreversibly, while entropy decrease requires significant effort. From this perspective, the iterative process of DRP is more natural, which is contrary to the approach of image restoration and artifact reduction. The entropy trajectory of the image is a prominent feature of DRP, which reveals the intrinsic quality of the framework and is instructively valuable for further research.

Fig. 6 and Fig. 7 show the convergence curves in the iteration process. The fluctuation of each epoch corresponds to the amplitude of the network parameter updates, and the overall trend is rhythmical and unidirectional.

*C. Method Comparison*

We utilize several images in CHAOS and Lung-PET-CT-Dx as test datasets to display the reconstruction results of various methods, which is showed in Fig. 10. To facilitate a reasonable visual comparison of the images, the image in Fig. 10 uniformly adopts the window width and window level of the grayscale. The display window is selected according to the histogram to ensure that the contrast is appropriate.

Visually, the resolution of the images reconstructed by DRP and FBPConvNet is nearly identical and surpasses that of DIP, ADMM and other algorithms. The red box in the image highlights the contrast in details. In some cases, DRP can even recover more details than FBPConvNet, which is an outstanding achievement for unsupervised algorithms. The graphical representations of statistical results of quantitative indicators are presented in Fig. 8. It is gratifying that the reconstructed images generated by DRP are of the highest quality in all cases. The blue box represents the indicator of DRP, which is obviously the best in each group of cases. On the other hand, although DRP outperforms FBPConvNet in both quantitative indicators and details restoration performance, it brings unsmooth structures into images, which limits the visual quality. To further demonstrate the ability of our method to preserve structures, the horizontal profile was plotted as a dashed blue line in Fig. 11. It is evident that the profile of DRP is most consistent with the reference image across edges and in approximately homogeneous regions.

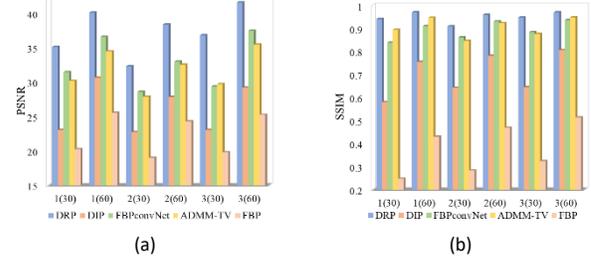

**Fig. 8.** Comparison among different methods for PSNR(a) and SSIM (b).

V. DISCUSSION

*A. Deep Radon Prior with single stage*

The above experiments indicate that the proposed iterative DRP framework is effective in sparse-view CT reconstruction, achieving superior PSNR/SSIM performance. However, the proposed framework has its deficiencies, one of which is that it is more computation cost-consuming than neural network-based algorithms with pre-trained models. DRP framework is essentially an iterative reconstruction, where neural network serves as a regularization function of the iteration and the input of neural network is changeable in every epoch. Thus, the pre-trained model of the neural network is unavailable. One of the feasible solutions to accelerate reconstruction progress is to adopt DRP framework with a single stage, which removes the iteration process and optimizes the reconstruction result by the neural network alone. In this case, the input of the network is fixed and the reconstruction framework is similar to the neural network-based method.

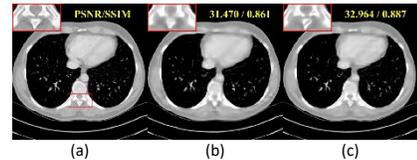

**Fig. 9.** Comparison among the results of DRP(c) and Single-stage DRP(b) under 30-views configuration to reconstruct the label (a). The display window is [-700, 400] HU. Comparison among the results of DRP(c) and Single-stage DRP(b) under 30-views configuration to reconstruct the label (a). The display window is [-700, 400] HU.

To validate the effectiveness of the DRP with single stage, we train a network with deep radon prior loss within 30 views. The training details is same as in section III-B, and the iteration number is 40000. Fig. 9 shows the reconstruction results of single stage DRP and the proposed DRP. The results indicate that the proposed iterative DRP framework is more accurate than single-stage DRP, which demonstrates the effectiveness of the proposed framework.

## B. Model Size and Complexity

The architecture and complexity of the neural network in the proposed DRP is also an essential factor to reconstruction result. Here, we further discuss the proposed DRP performance on neural networks with different complexity. A lighter model and a larger model with same architecture with Fig. 3 in section III-C are evaluated in this part. The channels of different layers in the network is denoted as [n1, n2, n3, n4, n5], while the encoder and decoder part are symmetric. The number of channels in light model is changed to [2, 4, 8, 16, 32] from [8, 16, 32, 64, 128], while that of the large model is [32, 64, 128, 256, 512].

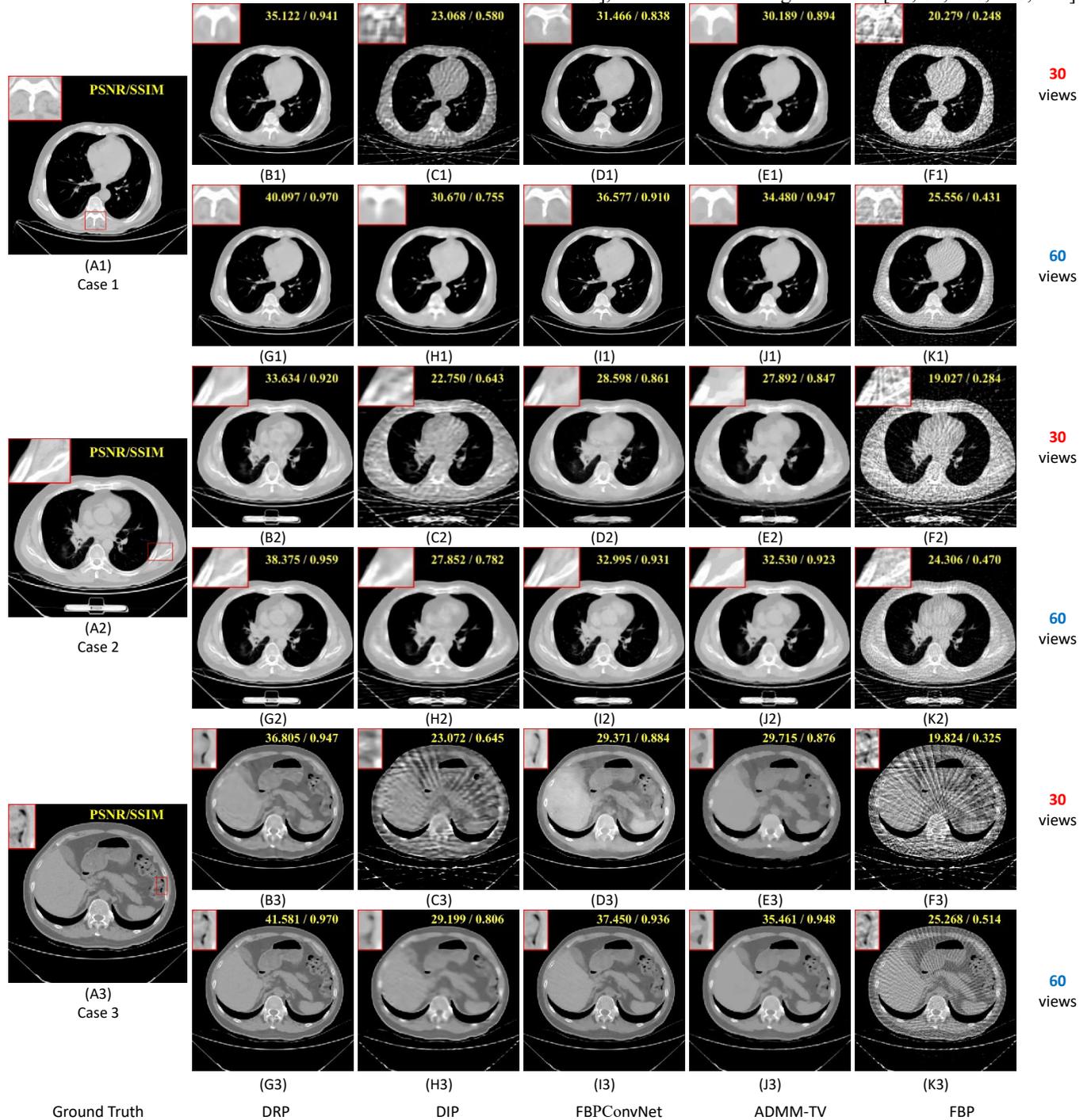

**Fig. 10.** The reconstructed images for different methods under sparse view. For each image, the first row is the results using 30 projections, while the second row is those in 60 projections. All of the image data is from two open source datasets: CHAOS and Lung-PET-CT-Dx. The display window of case 1 is [-900, 250] HU. The display window of case 2 is [-800, 300] HU. The display window of case 3 is [-450, 350] HU.

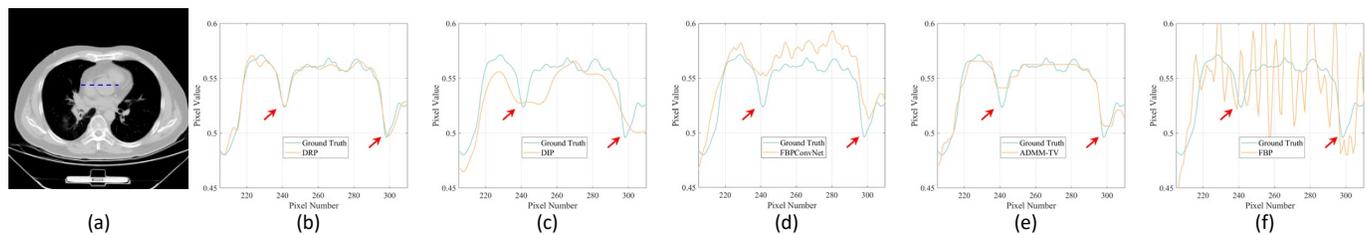

**Fig. 11.** The horizontal profiles along the dashed blue line in the reference image (a) versus the images reconstructed using (b) DRP, (c) DIP, (d) FBPConvNet, (e) ADMM-TV and (f) FBP respectively with 60 views. The red arrows indicate a selected region for the grayscale interval.

The reconstruction results of different model are shown in Fig. 12. Generally, a larger model is of greater representation ability, which leads to a better performance. As Fig. 12 (from (A1) to (D1)) shows, the increase of model complexity is helpful to improve the performance of DRP. However, in some cases, an oversized model is difficult to train and causes worse performance, as shown from Fig. 12 (A2) to Fig. 12(D2).

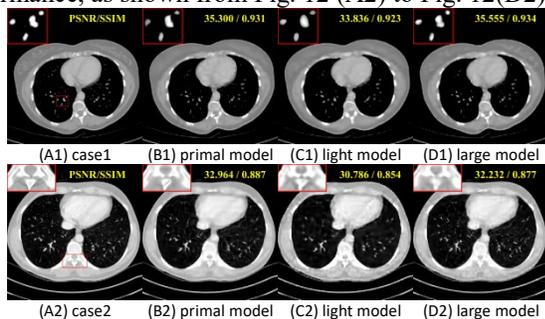

**Fig. 12.** The reconstruction images with 30 views for the reference images (A1) and (A2) using DRP with different model complexity from (B1) to (D1), and (B2) to (D2) respectively. The display window of the first row is [-600, 500] HU, while the display window of the second row is [-700, 400] HU.

The discussion of the aforementioned factors illustrates the effectiveness of the DRP framework and explores some limitations as well, which provides guidance for subsequent research. The reconstruction quality of the DRP framework can be further improved by optimizing the loss function, adjusting training strategies, and searching for network architecture and other hyperparameters.

## VI. Conclusion

In this work, we propose an unsupervised framework for sparse-view CT reconstruction called deep radon prior (DRP). The DRP framework combines neural network with traditional iterative reconstruction methods in the optimization goal, rather than functioning as an inverse operator. During the reconstruction process, the neural network is progressively optimized in multiple stages to narrow the solution space in the radon domain for the under-constrained imaging protocol. For the optimization strategy of DRP, this paper presents a set of feasible mathematical proofs under L2 norm, which provides a rigorous guarantee for the effectiveness of the framework and represents a breakthrough in terms of interpretability in unsupervised reconstruction algorithms. In comparison with popular supervised reconstruction methods, the proposed framework requires no training dataset while having better interpretability and generalization ability. The experimental results demonstrate that the proposed method achieves comparable or superior performance to the pre-trained methods. Moreover, the proposed method can be extended as a general optimization method to other ill-posed reconstruction problems, including limited angle CT, spectral CT, PET-CT, and other imaging technologies.